\documentclass[]{interact}
\usepackage{epstopdf}
\usepackage[caption=false]{subfig}
\usepackage{siunitx}
\usepackage{dcolumn}
\usepackage{hyperref}
\usepackage[numbers,sort&compress,merge]{natbib}
\bibpunct[, ]{[}{]}{,}{n}{,}{,}

\newcolumntype{d}[1]{D{.}{.}{#1}}

\theoremstyle{plain}

\theoremstyle{definition}

\theoremstyle{remark}

\newcommand{\mainAlFTrans}{A$^1\Pi \leftarrow $X$^1\Sigma^+$ }

\newcommand{\AlFgdstate}{X$^1\Sigma^+$ }

\newcommand{\AlAtomBrightnessLIF}{$\SI{3.0e12}{}$}
\newcommand{\AlAtomBrightnessAbs}{\SI{4.1e12}{}}

\newcommand{\AlFBrightnessLIF}{$\SI{1.6e12}{}$}
\newcommand{\AlFBrightnessAbs}{$\SI{2.1e12}{}$}

\newcommand{\CaAtomBrightnessLIF}{$\SI{9.2e12}{}$}
\newcommand{\CaAtomBrightnessAbs}{\SI{7.4e12}{}}

\newcommand{\CaFBrightnessLIF}{$\SI{1.4e11}{}$}
\newcommand{\CaFBrightnessAbs}{$\SI{1.1e11}{}$}

\newcommand{\MgFBrightnessLIF}{$\SI{1.1e11}{}$}
\newcommand{\MgFBrightnessAbs}{$\SI{1.1e11}{}$}

\newcommand{\YbAtomBrightnessAbs}{\SI{1.2e13}{}}
\newcommand{\YbAtomBrightnessLIF}{$\SI{5.0e12}{}$}

\newcommand{\YbFBrightnessLIF}{$\SI{0.7e11}{}$}
\newcommand{\YbFBrightnessAbs}{$\SI{2.1e11}{}$}

\newcommand\blfootnote[1]{%
  \begingroup
  \renewcommand\thefootnote{}\footnote{#1}%
  \addtocounter{footnote}{-1}%
  \endgroup
}

\begin{document}

\articletype{JOURNAL ARTICLE}

\title{Cryogenic Buffer Gas beams of AlF, CaF, MgF, YbF, Al, Ca, Yb and NO -- a comparison}

\author{
\name{Sidney C. Wright\textsuperscript{1,a}, Maximilian Doppelbauer\textsuperscript{1,a}, Simon Hofsäss\textsuperscript{1}, H. Christian Schewe\textsuperscript{1,*}, Boris Sartakov\textsuperscript{1}, Gerard Meijer\textsuperscript{1}, Stefan Truppe\textsuperscript{1,$\dagger$}}
\affil{\textsuperscript{1}Fritz Haber Institute of the Max Planck Society, Faradayweg 4-6, 14195 Berlin, Germany\\
\textsuperscript{*}Current Address: Institute of Organic Chemistry and Biochemistry of the Czech Academy of Sciences, Flemingovo nám. 2, 166 10 Prague 6, Czech Republic\\
\textsuperscript{$\dagger$}Current Address: Centre for Cold Matter, Blackett Laboratory, Imperial College London,
Prince Consort Road, London SW7 2AZ, United Kingdom
}
}

\maketitle

\begin{abstract}
Cryogenic buffer gas beams are central to many cold molecule experiments. Here, we use absorption and fluorescence spectroscopy to directly compare molecular beams of AlF, CaF, MgF, and YbF molecules, produced by chemical reaction of laser ablated atoms with fluorine rich reagents. The beam brightness for AlF is measured as $2\times 10^{12}$ molecules per steradian per pulse in a single rotational state, comparable to an Al atomic beam produced in the same setup. The CaF, MgF and YbF beams show an order of magnitude lower brightness than AlF, and far below the brightness of Ca and Yb beams. The addition of either NF$_3$ or SF$_6$ to the cell extinguishes the Al atomic beam, but has a minimal effect on the Ca and Yb beams. NF$_3$ reacts more efficiently than SF$_6$, as a significantly lower flow rate is required to maximise the molecule production, which is particularly beneficial for long-term stability of the AlF beam. We use NO as a proxy for the reactant gas as it can be optically detected. We demonstrate that a cold, rotationally pure NO beam can be generated by laser desorption, thereby gaining insight into the dynamics of the reactant gas inside the buffer gas cell.  
\end{abstract}


\label{ch:cellcharacterisation}

\section{Introduction}
\blfootnote{$^\mathrm{a}$ These authors contributed equally to this work.}
Cryogenic buffer gas cooling is a versatile technique to produce intense atomic and molecular beams with a low forward velocity \cite{Willey1988, Maxwell2005, Hutzler2012, Truppe2018} and allows efficient cooling of the rotational and vibrational degrees of freedom in molecules. Buffer gas cooling is now routinely used for precision spectroscopy and measurements \cite{Patterson2013, Collaboration2018, Alauze2021,Pilgram2021}, to trap molecules using magnetic fields \cite{DeCarvalho1999,Campbell2009,Lu2014} and electric fields \cite{Aggarwal2021}, to study collisions at low temperatures \cite{Skoff2011, Sawyer2011, Wu2017, Koller2022}, and to provide slow atoms and molecules for magneto-optical \cite{Hemmerling2014, Barry2014, Truppe2017, Anderegg2017, Collopy2018, Shaw2020, Lasner2021, Vilas2021} and electro-optical traps \cite{Prehn2016a}, in which the particles can be cooled to \SI{}{\micro\kelvin} temperatures. 

The species that have been cooled using a cryogenic buffer gas range from light atoms such as Li \cite{Singh2012} and K \cite{Lasner2021} to heavy atoms such as Yb \cite{Bulleid2013}, exotic atoms such as Tm, Er and Ho \cite{Hemmerling2014}, diatomic molecules \cite{Hutzler2011, Norrgard2017}, including radicals \cite{Maxwell2005, Lu2011, Bulleid2013, Hummon2013}, small polyatomic molecules \cite{Patterson2010b,Changala2016,Hutzler2020,Gantner2020} and large, complex molecules such as functionalised arenes \cite{Dickerson2021}, Nile Red \cite{Piskorski2014}, and C$_{60}$ \cite{BryanChangala2019}. 
Diatomic metal fluoride molecules with a $^2\Sigma^+$ electronic ground-state, such as MgF, CaF, SrF, BaF and YbF have attracted attention due to their suitability for direct laser cooling \cite{Shuman2009,Zhelyazkova2014,Yang2017b,Lim2018} and precision measurements \cite{Albrecht2020, Alauze2021}. Buffer gas molecular beams of these molecules have been reported and characterised \cite{Barry2012,Truppe2018,Albrecht2020,Skoff2011}. Recently, the molecules AlF and AlCl, which have $^1\Sigma^+$ electronic ground states, have been produced for the first time in buffer gas sources and are now the subject of laser cooling efforts \cite{Truppe2019a, Hofsass2021, Shaw2021, Daniel2021, Lewis2021}. These molecules are fundamentally different to the $^2\Sigma^+$ ground state molecules that have been laser cooled thus far, and we recently reported a bright molecular beam of AlF and fast optical cycling on its \mainAlFTrans transition \cite{Hofsass2021}.

It is often challenging to determine whether molecular yield in one experiment differs from another because of the choice of molecule, the detection method or conditions of the molecular beam source which may deteriorate over time. Here, we investigate beams of AlF, CaF, MgF, YbF, NO, Al, Ca and Yb in the same apparatus and use absorption and laser induced fluorescence spectroscopy to characterise them. By comparing the brightness of the molecular and atomic beams, we gain new insight in the production efficiency for the various species. We also describe observations which we consistently find produce desirable molecular beam properties and better long term performance.    
\section{Experimental Setup}
\subsection{Molecular Beam Source}

\begin{figure}
    \centering
    \includegraphics[width=\textwidth]{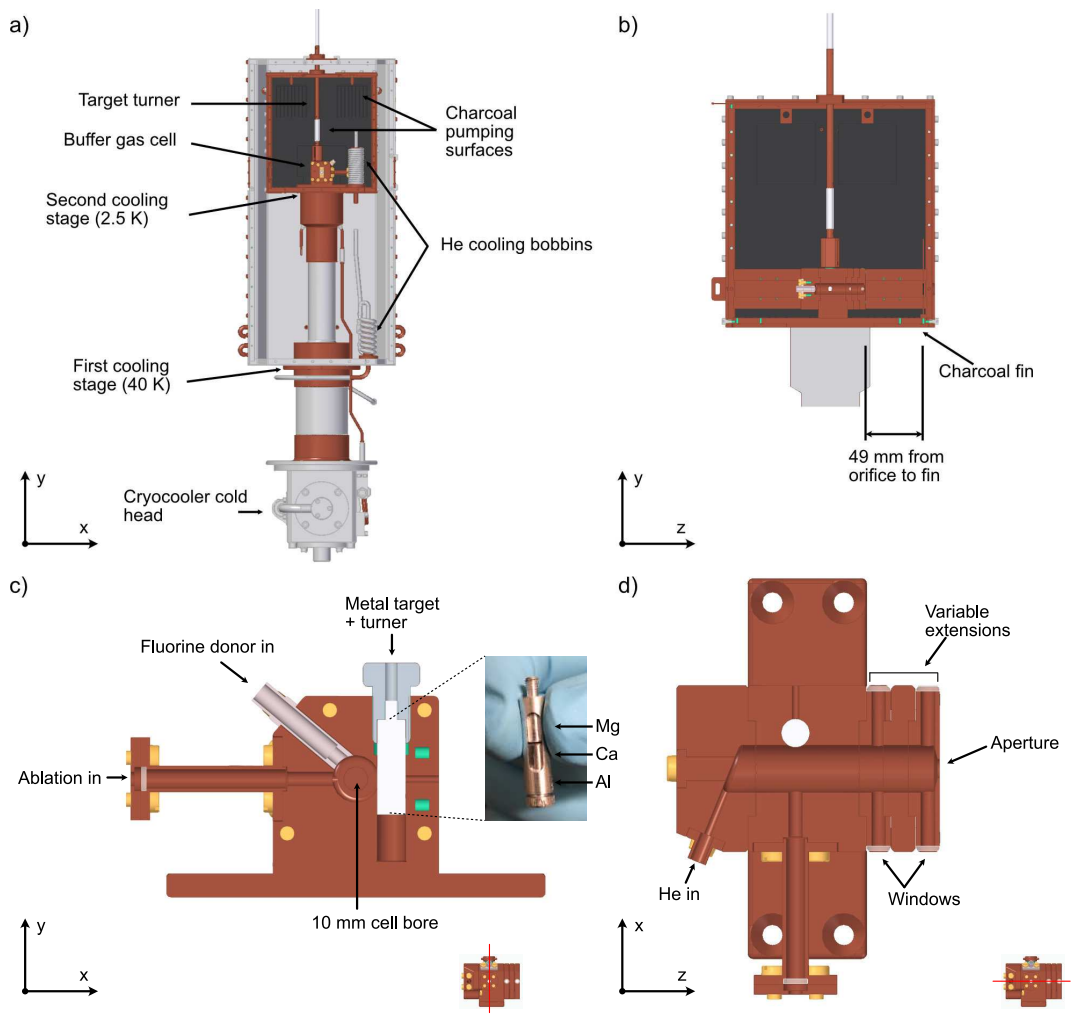} 
    \caption{(a): A diagram of the molecular beam source. The  molecular beam travels into the page. The front plates have been omitted to give a complete view of the inside of the source chamber. (b): Cut through the charcoal-coated shield along the direction of the molecular beam. (c): Cut through the buffer gas cell, perpendicular to the molecular beam direction, showing the region where the metal is ablated and reacts with a fluorine donor gas. The inset is a photograph of the multi-species metal target. (d): cut along the molecular beam direction showing the path of the He buffer gas.}
    \label{fig:exp:Exp_bothstages}
\end{figure}

Figure \ref{fig:exp:Exp_bothstages}a shows a sketch of the molecular beam source. We use a two-stage closed-cycle He cryocooler (Sumitomo RDK-415DP (with helium pot) with F-50H compressor) to cool the source to about \SI{2.5}{\kelvin}. The first stage cools the aluminium radiation shields to about \SI{40}{\kelvin}. The buffer gas cell, shown in Figure \ref{fig:exp:Exp_bothstages}c and d, is attached to the second stage and surrounded by a copper box, whose internal walls are coated with activated charcoal. The charcoal acts as an efficient sorption pump for helium gas at temperatures below \SI{10}{\kelvin}. The operating pressure of the source chamber under cryogenic conditions is about $1\times$10$^{-8}$~mbar, measured outside the radiation shields, and typically increases to 3$\times$10$^{-8}$~mbar when 2 standard cubic centimetres per minute (\SI{}{sccm}) of helium is flowed through the cell. 
To increase the effective pumping area, we attach charcoal-coated copper fins to the top internal face of the copper box. An additional single fin is mounted along the molecular beam direction (shown in Figure \ref{fig:exp:Exp_bothstages}b). While the top-fins have only a minor influence on the beam properties, the fin along the molecular beam helps maintaining a high molecular beam flux. Saturation of the activated charcoal with reaction products and ``ablation dust" --- indicated by a discoloration of the charcoal around the aperture --- can reduce the downstream flux by up to a factor of ten, without noticeably changing absorption outside the cell. In this case, the fin can be easily replaced. Increasing the distance between the cell and the first charcoal-coated surface mitigates this problem without affecting the downstream flux considerably. Cartridge heaters can be used to heat the set-up to room temperature within about five hours. 


A sketch of the buffer gas cell is shown in Figure \ref{fig:exp:Exp_bothstages}c and d. It is based on the design presented by Truppe \textit{et al.} \cite{Truppe2018}, with minor changes to facilitate machining and cleaning. The cell is machined from oxygen-free copper, has a circular bore with a 10~mm diameter, and its length is variable between 30 and \SI{60}{\milli\meter}, using extension pieces. Windows on the extensions allow in-cell absorption spectroscopy. For the experiments presented in this study, we used a \SI{40}{\milli\meter} long cell. The exit aperture of the cell has a diameter of $\SI{4}{mm}$. The multi-species metal ablation target is attached to a copper adaptor with a fine thread, which enables rotation of the target via a mechanical vacuum feed-through. We find that translating the target is superior to translating the ablation-laser beam and results in more consistent molecular beam properties. The relatively small bore size of the cell results in a short extraction time for molecules and thus short molecular pulses \cite{Truppe2017b, Truppe2018}, ideal for Stark deceleration and chirped laser slowing. The helium gas is pre-cooled in two copper pipe bobbins that are attached to each of the two cooling stages (see Figure \ref{fig:exp:Exp_bothstages}a). With a buffer gas flow rate of \SI{1}{sccm} the in-cell helium density is $7\times10^{14}\SI{}{\per\cubic\centi\meter}$, where we assumed an ideal vacuum conductance of the exit aperture.  

The output of a pulsed Nd:YAG laser (Continuum Minilite II) with a pulse energy of up to \SI{40}{\milli\joule} and a pulse-length of 5-\SI{7}{\nano\second} is gently focused to produce a \SI{0.7}{mm} diameter spot size on the target. The Nd:YAG fires with a repetition rate of \SI{1}{\hertz} for all measurements in this study, and firing the ablation light defines $t=0$ for each molecular pulse. A higher repetition rate leads to an increased heat load, resulting in an increase of the rotational temperature of the molecular beams. We ablate a metal target and introduce a fluorine donor gas (SF$_6$, NF$_3$, CF$_4$, XeF$_2$) into the cell through a copper capillary that is thermally insulated from the cell. We also use this capillary to inject NO molecules. The temperature of the capillary is kept at about \SI{120}{\kelvin}.

\begin{figure}
    \centering
    \includegraphics[width=0.5\textwidth]{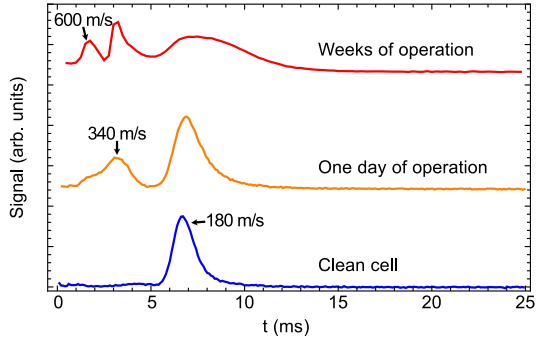}
    \caption{Fluorescence time of flight (TOF) traces of AlF with a buffer gas cell after different periods of use. From bottom to top: Cell freshly cleaned with Citranox detergent; same cell after one day of operation using the same ablation spot; same cell geometry after weeks of operation.}
    \label{fig:source:cleananddirty}
\end{figure}

We found over the course of our experiments that the beam properties of AlF are sensitive to the cell having clean internal surfaces. In particular, cleaning the cell in an ultrasonic bath with Citranox acidic detergent (Alconox, Inc.) results consistently in a slow  beam, with a small spread of arrival times seen in downstream fluorescence. With increasing number of ablation laser pulses, the time-of-flight broadens, and secondary higher velocity peaks appear. Removal of ablation products by wiping the inside of the cell leads to some recovery, but the efficacy of this method reduces gradually. Recovery of the original behaviour is reliably achieved by ultrasonic cleaning to expose the metal surface. We show examples of this behaviour for AlF in Figure \ref{fig:source:cleananddirty}. Coating the surfaces of the cell with a thin layer of gold to ensure a chemically inert surface was unsuccessful at preventing the observed degradation.

\begin{figure}
    \centering
    \includegraphics[width=\textwidth]{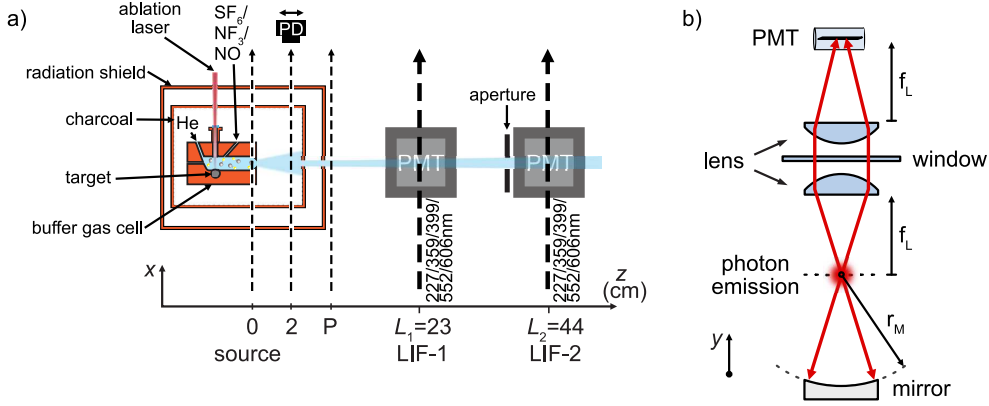}
    \caption{(a): Schematic view of the experimental setup used in this study to characterise the atomic and molecular beams. A photodiode (PD) is used for absorption and photomultiplier tubes (PMTs) for fluorescence spectroscopy. P marks the first downstream position where optical pumping can occur. Closer to the buffer gas cell collisions with helium redistribute the rotational state populations. (b): Optical layout of the fluorescence detector.}
    \label{fig:source:experimental_setup}
\end{figure}

A schematic of the full experimental setup used for this study is shown in Figure \ref{fig:source:experimental_setup}a. We use a multi-species target to quickly change the metal at the ablation spot, and therefore the species observed in the buffer gas beam. We can probe the atoms and molecules by absorption spectroscopy with a weak laser beam, both inside the cell and at a variable distance 0-\SI{2}{\centi\meter} from the exit aperture, detecting the transmitted light with a photodiode. The extraction efficiency as measured from the peak absorption of CaF, AlF and MgF inside and directly outside the cell is about $10-15\%$. The Doppler width measured inside and outside the cell is the same, but downstream of the cell, an increase of about a factor two in the transverse velocity spread is observed. We discuss the implications of this broadening for beam brightness measurements later.

Two laser-induced fluorescence (LIF) detectors enable us to measure the brightness of the collimated molecular beam, its velocity distribution and the rotational state distribution. The first detector, LIF-1, is located at $L_1=\SI{23}{\centi\meter}$ downstream of the cell exit. It is primarily used as a region for optical pumping, for instance to determine the velocity of the molecules using a pump-probe method \cite{Hofsass2021}. The second zone, LIF-2, is located $L_2=\SI{44}{\centi\meter}$ downstream of the cell exit and is used for spectroscopy and beam brightness measurements. The entrance to LIF-2 is restricted by a square aperture of size 2$\times$\SI{2}{\square\milli\meter} to reduce the transverse velocity spread of the molecular beam to about \SI{1.5}{\meter\per\second}. This reduces the Doppler broadening in our fluorescence excitation spectra and defines the interaction volume with the probe laser beam.

Figure \ref{fig:source:experimental_setup}b shows the collection optics of LIF-2. The fluorescence of the molecular beam is collimated by a \SI{50}{mm} plano-convex lens and concave mirror mounted in vacuum, and focussed onto a Hamamatsu R928 photomultiplier tube (PMT) by a second externally mounted lens. We collimate the excitation laser light to a $e^{-2}$ intensity diameter of 2-\SI{3}{\milli\meter}, to ensure that all molecules that enter the detection zone are addressed by the excitation laser light. The aperture of the detector subtends a solid angle $\delta\Omega=2.1\times10^{-5}$ \SI{}{\steradian} at the buffer gas cell exit and the peak optical power reaching the PMT when probing the brightest atomic beam is about \SI{1}{\nano\watt}. The PMT photocurrent is sent to a transimpedance amplifier with a current-to-voltage conversion factor of $10^5$ \SI{}{\volt\per\ampere}.

We use the same optical setup in the fluorescence detector for all species. The wavelengths of the fluorescence light span from 227 to \SI{606}{\nano\meter} and the collection efficiency of the LIF detector varies across this range due to differences in the PMT quantum efficiency and chromatic aberration of the detection optics. To compare the detection efficiency of the PMT across the range of emission wavelengths, we illuminated the PMT directly with about \SI{1}{\nano\watt} of laser light at 227, 360, 452 and 606\,nm. This was done by calibrating a $\sim 10^{-3}$ neutral density filter at incident powers of order \SI{1}{mW}, and then using the filter to attenuate to \SI{1}{\micro\watt}, measured using a calibrated optical power meter. To account for differences in collection efficiency, we use ray tracing simulations combined with the measured transmission and reflection losses of the optical components. 

\subsection{Molecular Beam Brightness Measurement}

The figure of merit for this study is the number of molecules produced per unit solid angle per pulse, in a given rovibrational level. To estimate this for the different species, we use both absorption and fluorescence measurements. 

Absorption measurements have the advantage of directly measuring the column density of the species of interest, but sufficient density is only available inside or near the exit of the buffer gas cell. The Lambert-Beer law relates the density, $n(t)$, to the measured absorption,

\begin{equation}
    -\ln(I_t/I_0) = n(t)\sigma(\nu) z, 
\end{equation}
\noindent Here, $\sigma(\nu)$ is the absorption cross-section, $z$ is the interaction length, and $I_t$ and $I_0$ are the intensity transmitted with and without the absorbing species present, respectively. In the absence of Doppler or hyperfine broadening, the maximum value of $\sigma(\nu)$ is the resonant absorption cross section $\sigma_{0}$, which for a transition $J'\leftarrow J$ is given by \cite{Budker2008}
\begin{equation}
    \sigma_{0}=\frac{\lambda^2}{2\pi}\frac{2J'+1}{2J+1}\frac{A_{ij}}{\Gamma} \hspace{0.1cm}.
\end{equation}
Here, $J$ and $J'$ are the ground and excited state total electronic angular momentum quantum numbers for atoms and the total electronic angular momentum including the end-over-end rotation for molecules, $A_{ij}$ is the Einstein A-coefficient of the resonant transition, and $\Gamma$ is the total spontaneous decay rate. In the presence of broadening mechanisms, the peak absorption cross-section $\sigma_D$ can be calculated using the relation $\int \sigma(\nu) d\nu = \sigma_0 \Gamma \frac{\pi}{2}$, where the integral over frequency $\nu$ covers the full absorption spectrum. 

We estimate the total number, $N$, by integrating the flux outside the cell over time, 

\begin{equation}
    N = v_f a_{aperture}\int n(t) dt =  \frac{v_f a_{aperture}}{z \sigma_D} \int -\ln(I_t/I_0) dt \hspace{0.1cm},
\end{equation}
\noindent where $a_{aperture}$ is the area of the exit aperture of the cell, $z$ is assumed to be the aperture diameter, and $v_f$ is the forward velocity which can be estimated from measurements downstream of the source. The brightness of the beam in absorption, $\mathcal{B}_{\mathrm{abs}}$, is then calculated as,

\begin{equation}
   \mathcal{B}_{\mathrm{abs}} = \frac{N}{\Delta\theta^2}
\end{equation}
\noindent with $\Delta\theta = 2\tan^{-1}(v_t/2v_f)$ the angular divergence of the beam, and $v_t$ the full-width at half maximum (FWHM) of the transverse velocity distribution, which is obtained from the absorption spectrum.

Fluorescence measurements, on the other hand, are much more sensitive and can be carried out far from the cell, but are typically difficult to calibrate on an absolute scale. For an optically closed transition, the number of photons scattered depends on the effective intensity and interaction time of the laser, and is thus sensitive to the laser beam profile and velocity of the species of interest. If the fluorescence can be saturated by optically pumping population to another state, a known number of photons $\mathcal{N}$ is absorbed and only the overall detection efficiency $\epsilon(\lambda)$ remains uncertain. When exciting from an initial state $i$ to an excited state $j$, $\mathcal{N}$ is given by,

\begin{equation}
    \mathcal{N}= \frac{\Gamma}{\Gamma - A_{ji}},
\end{equation}
\noindent where $A_{ji}$ is the Einstein-$A$ coefficient for decay back to the initial state and $\Gamma = \Sigma_n A_{jn}$ for all states $n$ is the total decay rate of the excited state $j$. If $N$ molecules enter the fluorescence detector, the number of photons detected is then simply $N\mathcal{N}\epsilon(\lambda)$. The on-axis brightness is then calculated as,

\begin{equation}
    \mathcal{B}_{\mathrm{f}} = \frac{N}{\delta\Omega} \hspace{0.1cm},
\end{equation}
\noindent This method can be easily applied to diatomic molecules with diagonal Franck-Condon factors, where it is straightforward to optically pump molecules using rotationally open electronic transitions, and $\mathcal{N}$ can be predicted with knowledge of the relevant Hönl-London factors. For atoms, optically open electronic transitions are commonly available, but often the coefficients $A_{ji}$ are not known accurately.

\subsection{Laser Systems}
We use two different continuous laser systems for this study. To detect Al, Ca, Yb, AlF, MgF and NO, we use a Ti:Sapphire laser (MSquared Solstis), whose output is frequency doubled in successive enhancement resonators containing a nonlinear optical crystal. The linewidth of the fundamental light is less than 400\,kHz. A single stage of frequency doubling is sufficient to generate light near \SI{360}{\nano\meter} (MgF) and \SI{399}{\nano\meter} (Yb), and we use two successive doubling stages to generate UV light near \SI{227}{\nano\meter} (AlF, NO, Al, Ca). The second laser system is a Coherent 899 ring dye laser (RDL) which generates light near \SI{606}{\nano\meter} for detection of CaF and near \SI{552}{\nano\meter} for the detection of YbF. The laser linewidth is around \SI{1}{MHz}. The laser frequencies are monitored with a HighFinesse WS8-10 wavemeter calibrated using a temperature-stabilised HeNe laser. Details on the absolute accuracy of the wavemeter can be found in reference \cite{Doppelbauer2021a}\footnote{Probing the Mg atoms would require light at 285\,nm to excite the $^1P_1 \leftarrow {}^1S_0$ transition, which was not available for this study.}. 

\section{Atomic beam measurements}
We first discuss measurements on the atomic buffer gas beams of Ca, Al, and Yb. These provide an important reference when comparing with the molecular beams.
\subsection{Aluminium}
To characterise the Al atomic beam, we use the $(3s^25d)^2D_{3/2}\leftarrow(3s^23p)^2P_{1/2}$ transition near \SI{226.4}{\nano\meter}, conveniently close to the \mainAlFTrans transition in AlF. Due to the fine structure splitting of the electronic ground state, there exists a $(3s^23p)^2P_{3/2}$ state approximately \SI{112}{\per\centi\meter} ($\sim 160$\,K) above the ground state. By probing the nearby $(3s^25d)^2D_{5/2}\leftarrow(3s^23p)^2P_{3/2}$ transition, we find that about 1\% of the Al atoms exiting the source initially occupy the $^2P_{3/2}$ state, and they appear only with high velocities greater than \SI{500}{\metre\per\second}. We searched for evidence of Al$_2$ dimers by probing the  E $^3\Sigma_g^-\leftarrow $X$^3\Pi_{0u} $ transition near \SI{366.6}{nm} \cite{Fu1990}. We therefore assume the population in the $(3s^23p)^2P_{1/2}$ state represents the full atomic population.

\begin{figure}
    \centering
    \includegraphics[width=\textwidth]{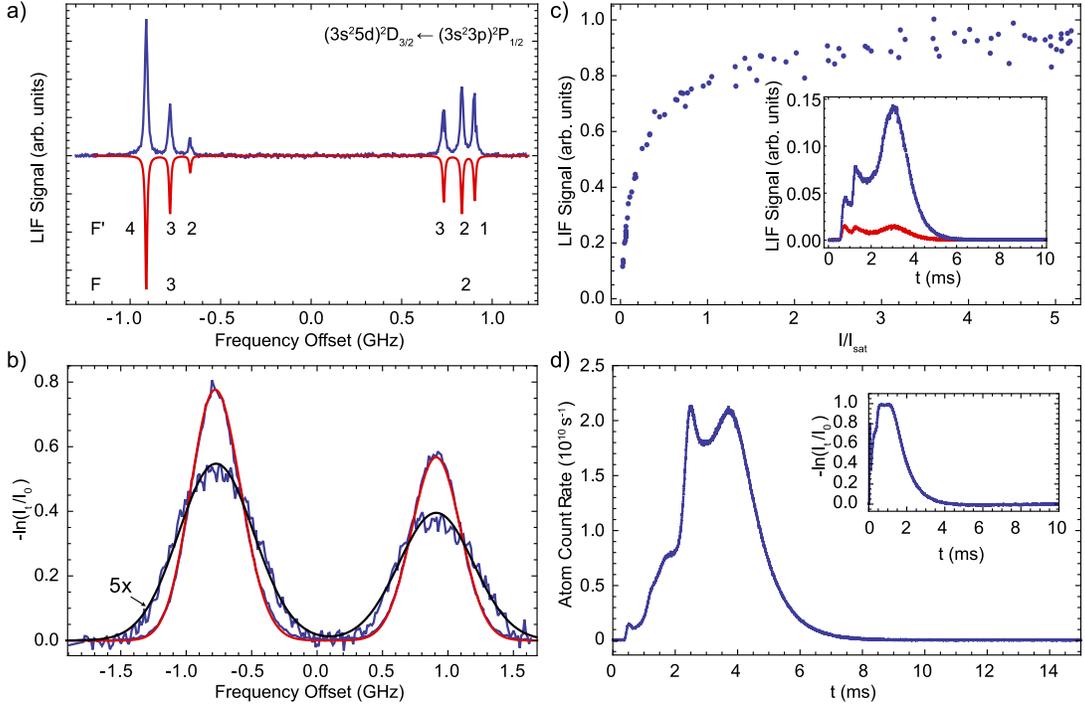}
    \caption{(a): Hyperfine-resolved LIF spectrum of the $(3s^25d)^2D_{3/2}\leftarrow (3s^23p)^2P_{1/2}$ transition of Al near 227.6 nm. The experimental spectrum is shown in blue. A simulated spectrum using equations \ref{eqn:AlHFenergies} and \ref{eqn:AlLineStrengths} and the known linewidth of $12.8$\,MHz is shown in red. The ground-state ($F$) and excited-state ($F'$) hyperfine quantum numbers of the transitions are given below the simulated spectrum. (b): Aluminium absorption spectra recorded directly outside the cell exit and \SI{10}{\milli\meter} downstream (magnified by a factor five). The solid curves in red and black are fits to the data (blue) using the known spectroscopic constants and a Gaussian lineshape to determine the Doppler broadening. The Doppler broadening increases for the downstream position. (c): Fluorescence saturation curve of the $(3s^25d)^2D_{3/2},F'=4\leftarrow(3s^23p)^2P_{1/2},F=3$ line of Al, demonstrating that we can saturate the fluorescence by optical pumping. The inset shows the LIF TOF without (blue) and with (red) a pumping laser on the same transition in LIF-1. (d): Time of flight traces of the Al beam under saturated fluorescence of the $(3s^25d)^2D_{3/2},F'=4\leftarrow(3s^23p)^2P_{1/2},F=3$ transition to determine the total flux through a $2\times2$ \SI{}{\square\milli\meter} aperture \SI{440}{\milli\meter} from the source aperture. The inset shows the absorption trace directly outside the cell.}
    \label{fig:source:Al_spectra}
\end{figure}

\begin{figure}
    \centering
    \includegraphics[width=0.5\textwidth]{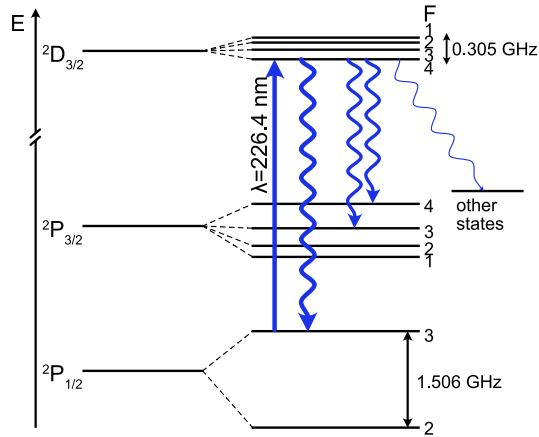}
    \caption{Scheme of the relevant electronic states and hyperfine levels of Al. Electronic energies are not to scale, the relative hyperfine splittings are shown to scale. We use the $^2$D$_{3/2},F'=4\leftarrow^2$P$_{1/2},F=3$ transition to detect Al atoms.}
    \label{fig:source:Al_levelscheme}
\end{figure}

The lifetime of the $(3s^25d)^2D_{3/2}$ excited state, $\tau_{\mathrm{ex}}$ is \SI{12.3}{\nano\second} ($\Gamma/2\pi=\SI{12.8}{\mega\hertz}$) \cite{Kelleher2008}, but to our knowledge the hyperfine structure of the $(3s^25d)^2D_{3/2}\leftarrow(3s^23p)^2P_{1/2}$ transition has not been measured. The nuclear spin $\mathbf{I} (I_{Al} = 5/2)$, couples with the electronic total angular momentum $\mathbf{J}$ to give eigenstates labelled by the total angular momentum quantum number $F$. Figure \ref{fig:source:Al_spectra}a shows a hyperfine-resolved experimental spectrum of the $(3s^25d)^2D_{3/2}\leftarrow(3s^23p)^2P_{1/2}$ transition, obtained using about \SI{100}{\micro\watt} of linearly polarised laser light. The larger splitting $\sim 1.5$\,GHz arises from the known ground state hyperfine interaction, and the interaction in the excited state leads to a further splitting of a few hundred MHz. The hyperfine levels are shown in Figure \ref{fig:source:Al_levelscheme}. We fit the hyperfine energies, $E_i(J,I,F)$, to the equation \cite{Davis1949}

\begin{equation}
   E_i =  \frac{A_i C}{2} + B_i\frac{\frac{3}{4}C(C+1)- I(I+1)J(J+1)}{2I(2I-1)J(2J-1)},
   \label{eqn:AlHFenergies}
\end{equation}
where $A_i$ is the interaction strength between the nuclear spin $\mathbf{I}$ and the electronic angular momentum $\mathbf{J}$, $B_i$ is the quadrupole interaction coefficient, necessarily zero for $J< 1$, and $C=F(F+1)-I(I+1)-J(J+1)$. The index $i$ labels the electronic state. For the ground state we find $A= 501.6(1.2)$\,MHz, consistent with the literature value \cite{Chang1990}. For the excited state, we determined $A=-33.9(4)$\,MHz and $B=1.9(2.5)$\,MHz.
Shown pointing downwards in red in Figure \ref{fig:source:Al_spectra} is a simulated spectrum using the known natural linewidth, where we assumed equal population of the ground state Zeeman sub-levels and isotropic emission of the fluorescence light. Under this assumption, the relative intensities of the $F',F$ hyperfine lines $l_{F',F}$ are given by,

\begin{equation}
    l_{F',F} =
    |\sqrt{(2F'+1)(2F+1)}
    \times
    \begin{Bmatrix}
    J' & F' & I \\
    F & J & 1 \\
    \end{Bmatrix}|^2 \hspace{0.1cm}.
    \label{eqn:AlLineStrengths}
\end{equation}

\noindent This formula describes the data well. The fitted FWHM of the lines is found as \SI{19}{MHz}, slightly larger than the measured linewidth, which we attribute to Doppler broadening, residual Zeeman shifts due to a small ambient magnetic field ($\sim 1$\,G) in the detection region, and the effect of optical pumping.     

The measured hyperfine structure can then be used to determine the Doppler broadening present in absorption measurements. Figure \ref{fig:source:Al_spectra}b shows two such absorption spectra, directly outside the exit of the buffer gas cell and \SI{10}{\milli\meter} further downstream. There and in the following absorption plots, we show the optical depth -$\mathrm{ln}(I_{\mathrm{t}}/I_0)$ on the $y
$-axis.
We fit the data using the measured excited state hyperfine structure convolved with a Doppler broadening term with FWHM $\Gamma_D$. The transverse velocity width, $v_t=\Gamma_D\lambda/(2\pi)$, is approximately \SI{90}{\meter\per\second} at $z=\SI{0}{\milli\meter}$ and \SI{150}{\meter\per\second} at $z=\SI{10}{\milli\meter}$. This increase of about a factor 1.7 indicates collisions of the Al atoms with (somewhat hotter) He buffer gas outside the cell, increasing the divergence of the beam. Using the downstream value for the transverse velocity spread, we estimate the brightness of the atomic beam as \AlAtomBrightnessAbs{}. 

To measure the Al beam brightness in fluorescence, we use the scheme shown in Figure \ref{fig:source:Al_levelscheme}. We excite the $F'=4\leftarrow F=3$ hyperfine line, which optically pumps 95\% of the population to the $(3s^23p)^2P_{3/2}$ state. The remaining 5\% of the atomic population is pumped to higher lying $^2P$ and $^2F$ electronic states, which cannot directly decay back to the electronic ground state by dipole allowed transitions. We neglect the effect of these states other than the fact that they slightly increase the optical pumping probability. From the ratio of the Einstein $A$-coefficients of these transitions, we calculate that we optically pump population in the $F=3$ hyperfine level after absorbing on average $6.1(1.1)$ photons. In Figure \ref{fig:source:Al_spectra}c, the fluorescence signal on this transition in LIF-2 is plotted against the laser intensity expressed in terms of the two-level saturation intensity $I_{\textrm{sat}}=\pi h c\Gamma/(3\lambda^3)$, where $\Gamma$ is the spontaneous decay rate and $\lambda$ is the transition wavelength. The fluorescence saturates, demonstrating that we can use this transition to accurately determine the average number of photons emitted by the atoms. The inset shows the change in the fluorescence signal when the same laser light is used in LIF-1 to optically pump the atoms before detecting the remaining atoms in LIF-2. The pumping efficiency is approximately 90\%, increasing with arrival time at the detector. After accounting for the $F=2$ ground state population which is not detected, we arrive at a total brightness of Al atoms of \AlAtomBrightnessLIF{}$\SI{}{\per\steradian}$ per pulse. This is within $25\%$ of the number derived from absorption measurements in the source when we assume the transverse velocity as measured \SI{10}{\milli\meter} downstream from the source. If, instead, we used the transverse velocity spread as measured right outside the cell, we would overestimate the beam brightness by a factor of 3.5. This clearly shows that absorption measurements close to the cell can be misleading, and typically will provide an upper bound to the beam brightness observed downstream.


\subsection{Calcium}

\begin{figure}
    \centering
    \includegraphics[width=0.5\textwidth]{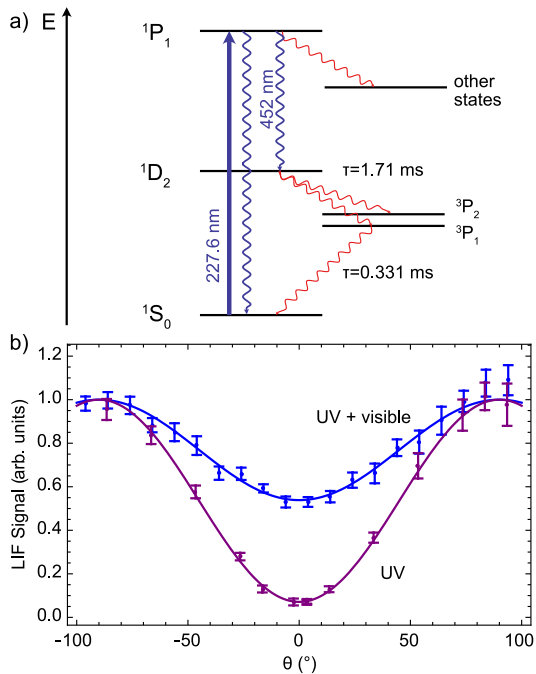}
    \caption{(a): Scheme of the relevant energy levels of calcium. (b): Polarisation dependence of fluorescence of the Ca atomic beam. The UV fluorescence, isolated with a 227\,nm bandpass filter, shows strong anisotropy due to the dipole emission pattern. The combined fluorescence in the UV and visible shows reduced emission anisotropy as the visible light emission is nearly isotropic.}
    \label{fig:source:Calevelscheme}
\end{figure}

To detect calcium, we use the $(3p^6 4snp) ^1P_1 \leftarrow (3p^64s^2) ^1S_0$ transition 
at \SI{227.6}{\nano\meter}. In Figure \ref{fig:source:Calevelscheme}a, the relevant energy levels for our study of the Ca atomic beam are shown. Exciting the atoms to the $^1P_1$ state primarily pumps the population to the metastable $(3p^6 3d 4s)^1D_2$ state, with emission of a $452$\,nm photon. There are additional decay channels to higher lying $^1S$ and $^2D$ states, but these emit at wavelengths longer than $940$\,nm, outside the detection range of the PMT. The direct decay channel back to the $^1P_0$ state emits a photon at 227.6\,nm. The Einstein-$A$ coefficient for this decay channel has been measured as $A(^1S_0)= 2.84(39)\times10^7$\SI{}{\per\second} \cite{Parkinson1976}. We collect both $227$\,nm and $452$\,nm photons and account for the detector sensitivity at the different wavelengths. The two decay paths have different fluorescence emission patterns, which is demonstrated by monitoring the LIF as a function of the angle of linearly polarised excitation light, $\theta$. Figure \ref{fig:source:Calevelscheme}b shows the detected fluorescence against $\theta$ with and without a filter blocking the visible light covering the PMT. The emission pattern of the $227$\,nm decay is proportional to $\sin^2\theta$, whereas the $452$\,nm emission is proportional to $1+\frac{1}{6}\sin^2\theta$, and almost isotropic. This leads to the reduced contrast in the fluorescence as a function of $\theta$ when detecting both the visible and UV fluorescence compared to the UV alone. From these measurements and the detector efficiencies, we estimated that the ratio of Einstein A-coefficients of the two transitions, $A(^1D_2)/A(^1S_0) = 2\pm0.3$. Using the measured value for $A(^1S_0)$, we find a total decay rate of $A(^1S_0) + A(^1D_2) = 8.5(1.4) \times 10^7$\SI{}{\per\second}, larger than the measured spontaneous emission rate of $\Gamma = 1/\tau = 6.37\times10^7$\SI{}{\per\second} \cite{DenHartog2021}. We assume that atoms in the Ca beam emit on average $\mathcal{N}_{227} = 0.5$ and $\mathcal{N}_{452} = 1$ photons at the two detection wavelengths when the fluorescence is saturated, but note that these values are approximate.

\begin{figure}
    \centering
    \includegraphics[width=\textwidth]{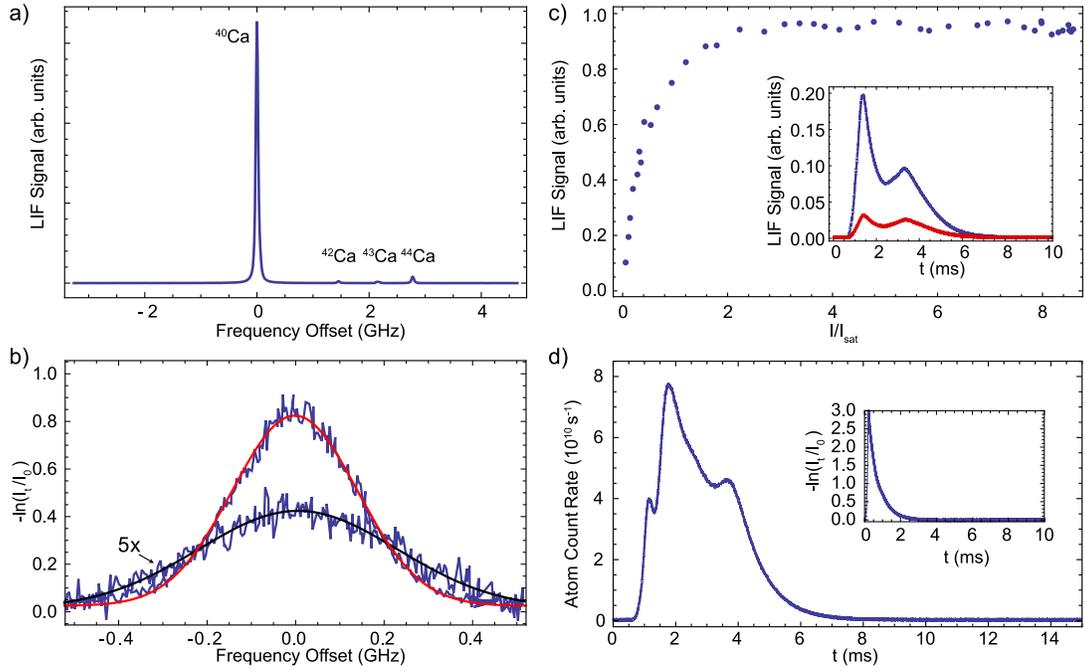}
    \caption{(a): LIF spectrum of the $(3p^6 4snp) ^1P_1 \leftarrow (3p^64s^2) ^1 S_0$ line of Ca. The observed isotope peaks are labelled. (b): Absorption spectra of the $(3p^6 4snp) ^1P_1 \leftarrow (3p^64s^2) ^1 S_0$ transition in $^{40}$Ca, recorded directly outside the cell exit (Gaussian fit shown in red) and \SI{20}{\milli\meter} downstream (black). (c): Fluorescence saturation curve of the $(3p^6 4snp) ^1P_1 \leftarrow (3p^64s^2) ^1 S_0$ transition of Ca. The inset shows the LIF TOF without (blue) and with (red) a pumping laser tuned the same transition frequency. (d): Fluorescence and absorption TOF traces of the Ca beam, using light resonant with the $(3p^6 4snp) ^1P_1 \leftarrow (3p^64s^2) ^1 S_0$ transition of $^{40}$Ca.}
    \label{fig:source:Califlevelscheme}
\end{figure}

The fluorescence spectrum is shown in Figure \ref{fig:source:Califlevelscheme}a, where we clearly resolve four naturally abundant Ca isotopes. The transitions of the two low-abundance bosonic isotopes, $^{42}$Ca and $^{44}$Ca, are shifted relative to $^{40}$Ca by $+1.43(1)$\,GHz and $+2.755(8)$\,GHz respectively. We hereon focus on $^{40}$Ca, which is relevant for comparison with $^{40}$CaF.

In Figure \ref{fig:source:Califlevelscheme}b, absorption spectra of the most abundant isotope peak of the $(3p^6 4snp) ^1P_1 \leftarrow (3p^64s^2) ^1 S_0$ transition are shown directly at the cell orifice and \SI{20}{\milli\meter} downstream. We again observe an increase in the transverse velocity width at the downstream position, from \SI{75}{\meter\per\second} to \SI{120}{\meter\per\second}.

In Figure \ref{fig:source:Califlevelscheme}c, the fluorescence signal in LIF-2 is plotted against peak laser intensity, demonstrating saturation of the fluorescence. The inset to the figure shows a pump-probe measurement, from which it appears that we are only able to pump $70\%$ of the atoms. However, the finite lifetime of the $^1D_2$ state of \SI{1.71}{\milli\second} \cite{Mills2017} is comparable to the flight time between the detectors. Partial repopulation of the ground state then occurs via the $^3P_1$ state, which is populated with a probability of 83\% and decays to the $^1S_0$ state with a lifetime of \SI{0.331}{\milli\second}. Consistent with this, the pumping efficiency increases for earlier arrival times at the detector, where the flight time between pump and probe is shorter. We therefore conclude that the population can be fully optically pumped for the purposes of measuring the beam brightness.

Figure \ref{fig:source:Califlevelscheme}d shows a time of flight trace of the Ca atoms in LIF-2. We calculate that $1.9\times10^8$ atoms pass through the detector, corresponding to an on-axis brightness of $9.2\times10^{12}$~sr$^{-1}$ per pulse. Using the absorption measurement 
(shown in the inset of Figure \ref{fig:source:Califlevelscheme}d), we find \CaAtomBrightnessAbs{} per pulse, in reasonable agreement. We conclude that the Ca atomic beam is brighter than the Al atomic beam, by about a factor 2.5.

\subsection{Ytterbium}
\begin{figure}
    \centering
    \includegraphics[width=0.5\textwidth]{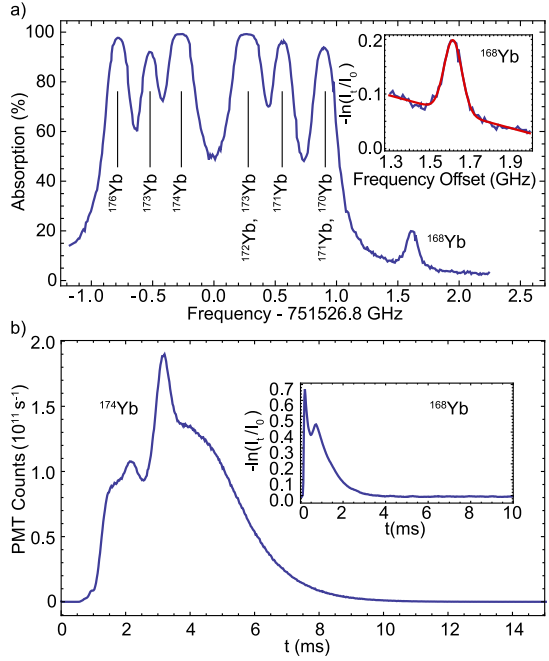}
    \caption{(a): Absorption spectrum of the $(6s6p)^1P_1\leftarrow(6s^2)^1S_0$ transition, for all stable isotopes of ytterbium. The inset shows a zoom-in on the $^{168}$Yb peak. The red line is a fit to determine the Doppler width against the background absorption of more abundant isotopes. (b): Fluorescence TOF traces of the ytterbium beam, using light resonant with the $(6s6p)^1P_1\leftarrow(6s^2)^1S_0$ transition for the $^{174}$Yb beam. Because the transition is optically closed, we show the PMT count rate which is proportional to the atomic density inside the detection volume. The inset shows an absorption trace for the $^{168}$Yb isotope taken directly outside the cell exit.}
    \label{fig:source:Ybabsspectra}
\end{figure}
We use the $(6s6p)^1P_1\leftarrow(6s^2)^1S_0$ transition near \SI{399}{\nano\meter} to study the Yb atomic beam. The excited state lifetime $\tau_{\mathrm{Yb}}=\SI{5.464}{\nano\second}$ $(\Gamma/2\pi=\SI{29.13}{\mega\hertz})$ \cite{Takasu2004}. 
An absorption spectrum taken outside the cell is shown in Figure \ref{fig:source:Ybabsspectra}a. Ablation of Yb metal is highly efficient and the absorption saturates for the most abundant isotopes. We are able to measure the absorption of the $^{168}$Yb isotope (Figure \ref{fig:source:Ybabsspectra}a, inset), and use this to estimate the number of atoms and to extract the Doppler width of the beam. The transverse velocity width FWHM at the cell orifice is \SI{42}{\meter\per\second}, and \SI{60}{\meter\per\second} when measured \SI{18}{\milli\meter} downstream. After taking into account the relative natural abundances of $^{168}$Yb$:^{174}$Yb (1:245), we estimate the brightness of the $^{174}$Yb isotope in absorption is $1.2\times10^{13}$~sr$^{-1}$ per pulse.

Figure \ref{fig:source:Ybabsspectra}b shows a time-of-flight profile obtained when probing the $^{174}$Yb isotope with $\SI{40}{\micro\watt}$ of laser power, to be compared later with the $^{174}$YbF molecule. The $(6s6p)^1P_1\leftarrow(6s^2)^1S_0$ transition is optically closed, and therefore not suitable for precise fluorescence measurements of the beam brightness. We simply estimate the brightness observed downstream by modelling the interaction of atoms with the mean forward velocity with the intensity profile of the probe light. Assuming a velocity of \SI{150}{\meter\per\second}, we estimate the beam brightness as $5.0\times10^{12}$~sr$^{-1}$ per pulse, in LIF-2, in reasonable agreement with the absorption measurements. We note that the total brightness considering all isotopes is at least $1.6\times10^{13}$ per steradian per pulse, making the Yb beam the brightest of the three atomic species tested.

\subsection{Addition of Fluorine Donor Gases}

\begin{figure}
    \centering
    \includegraphics[width=\textwidth]{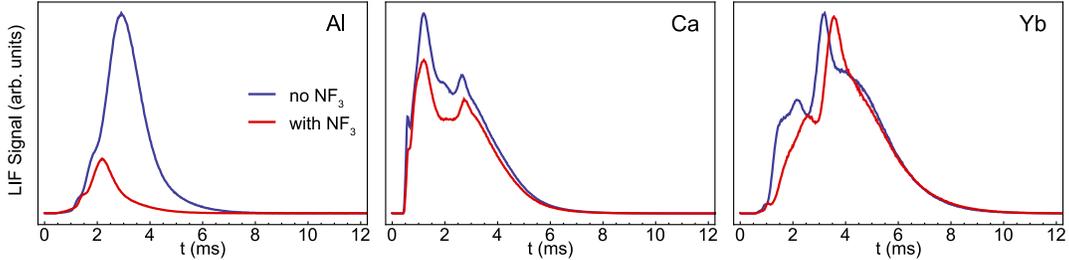}
    \caption{LIF signal traces of the Al, Ca and Yb atom beams without (blue) and with (red) NF$_3$ flow into the buffer gas cell, demonstrating the strong effect of a fluorine donor gas on the number of aluminium atoms in the beam.}
    \label{fig:source:AlCaYbwithwithout}
\end{figure}

Figure \ref{fig:source:AlCaYbwithwithout} shows the fluorescence TOF profiles of the atomic beams before and directly after introducing a flow of \SI{0.002}{sccm} NF$_3$ to the cell. For the Al atoms, there is an evident loss in signal, especially for the atoms with late arrival times in the detector. We typically observe a reduction of around 80\% in the total number of atoms reaching the detector in LIF-2. For Ca and Yb, the effect of the reactant gas is much less pronounced. We observe a 20\% and 10\% reduction in the atomic signal, respectively and obtain qualitatively similar results when using SF$_6$ as the reactant gas. However, the atomic beam speeds up upon excess addition of SF$_6$; we discuss observable effects on the molecular beam thermalisation later. Excited state chemistry can increase the yield of the species of interest in the source. This has been demonstrated for the reaction of Yb and Ca with water and alcohols to form YbOH and CaOH \cite{Oberlander1996, Jadbabaie2020, Vilas2021} and for the reaction of carbon atoms with hydrogen molecules to form CH \cite{Ikejiri2005}. We found no influence on CaF production when exciting the Ca atoms with \SI{40}{mW\per\centi\metre\squared} of $(3p^6 4snp) ^1P_1 \leftarrow (3p^64s^2) ^1 S_0$ light longitudinally through the cell. 

The exact process through which the reaction between the atoms and the reactant gas happens is largely unknown, but available experimental data provides some useful information. The activation energies of the reactions Al $+$ NF$_3$ $\rightarrow$ AlF $+$ NF$_2$ and Al $+$ SF$_6$ $\rightarrow$ AlF $+$ SF$_5$ have been experimentally measured as 2990~K and 4800~K respectively \cite{Parker2002}. This can be compared to the kinetic energy of the ablated Al atoms, which decreases exponentially with the number of Al-He collisions \cite{Hutzler2012}. Measurements of the Doppler width of the Al atoms in the absence of the buffer gas suggested an initial temperature of $3400\pm\SI{1000}{\kelvin}$. Even assuming an initial Al temperature of $10^4$\,K, the typical energy for the reaction falls below the activation energy after 4 (SF$_6$) and 7 (NF$_3$) Al-He collisions. This suggests that for a two-body gas phase reaction, the reaction must be highly efficient to generate an appreciable number of molecules.

A number of observations suggest that the reactant gas in fact remains frozen in the cell, and is brought into the gas phase in the ablation process. Firstly, removing the flow of reactant gas into the cell with a tap results in a slow drop of the molecular beam signal, over 200 shots, but never fully to zero. However, this requires active ablation of the target; if both the ablation light and the reactant gas flow are removed for several hundred shots, and the ablation light is then introduced again, the signal decay begins from its original value. We have often observed a sizeable fraction of signal for $\sim 30$ minutes of continued operation with the reactant gas tap completely shut. 

\begin{figure}
    \centering
    \includegraphics[width=\textwidth]{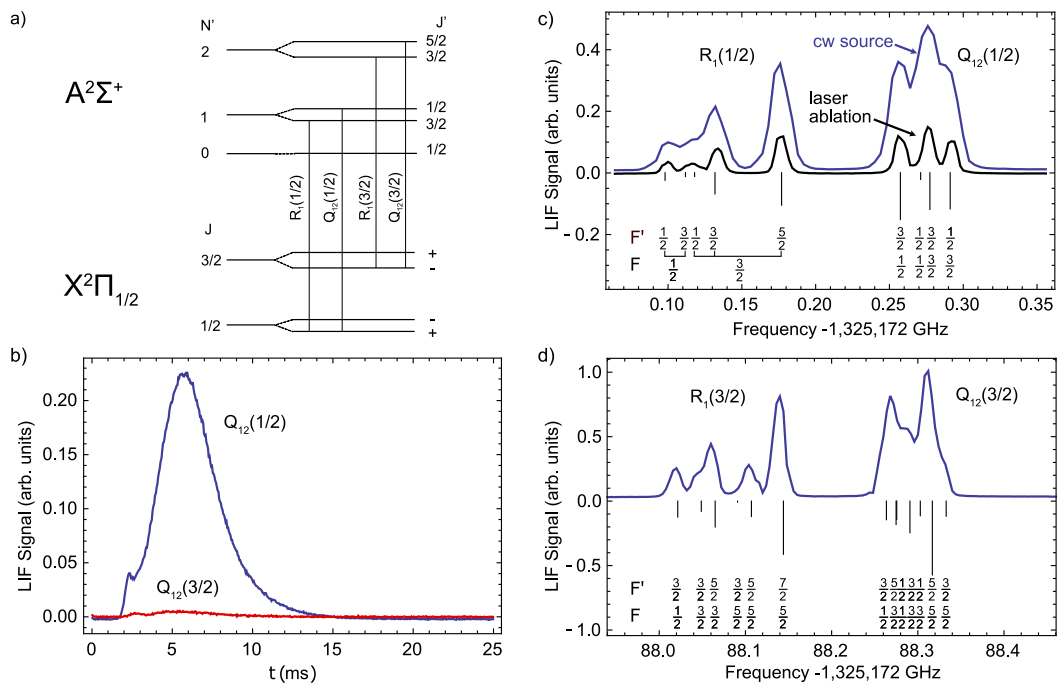}
    \caption{Experiments with nitric oxide (NO) flowed into the buffer gas cell through the fluorine donor gas inlet. (a): Scheme of the relevant energy levels (hyperfine structure omitted). (b): Firing the ablation laser generates a pulse of rotationally cold, slow NO molecules. (c): a spectrum of the R$_1$(1/2) and Q$_{12}$(1/2) lines taken using the pulsed desorption method with the buffer gas cell at 2.5\,K, and a spectrum using a continuous beam with the buffer gas cell at 70\,K. (d): A spectrum of the R$_1$(3/2) and Q$_{12}$(3/2) lines taken using the continuous source shown on the same vertical scale as (c).}
    \label{fig:NO}
\end{figure}

To investigate this further, we replaced the flow of reactant gas with \SI{0.01}{sccm} of nitric oxide (NO), and fired the ablation laser at the metal target to reproduce the usual conditions in the cell. NO can be optically detected via the A$^2\Sigma^+,v'=0\leftarrow$ X$^2\Pi_{1/2},v=0$ transition near 226.2\,nm, and its freezing point (109\,K) is intermediate between SF$_6$ (223\,K) and NF$_3$ (66\,K). The hyperfine structure of this transition has been resolved in a number of studies \cite{Engleman1969,Meerts1972,Gray1993,Reid1994,Brouard2012}, and absorption spectroscopy in a room-temperature cell was recently used to measure hyperfine structure of high-$J$ lines of the A$^2\Sigma^+,v'=0\leftarrow$ X$^2\Pi_{3/2},v=0$ transition \cite{Kaspar2022}. Here, we probe the two lowest energy levels of the X$^2\Pi_{1/2},v=0$, clearly resolving the hyperfine structure. A level scheme for NO is shown in Figure \ref{fig:NO}a. The Franck-Condon matrix for this transition is highly non-diagonal \cite{Engleman1969}, and to a good approximation we can assume a single photon absorption will optically pump an NO molecule to other rovibrational levels of the X$^2\Pi$ state. 

With the buffer gas cell cooled to $2.5$\,K, we observe a beam of buffer gas cooled, ground-state NO molecules, which we assume are desorbed upon firing the ablation laser (Figure \ref{fig:NO}b). The mean forward velocity of this beam is about \SI{120}{\metre\per\second} and we estimate the brightness as $10^{12}$ molecules per steradian per pulse, with a stability of better than $10\%$ (standard deviation). The signal in the $J=3/2$ first rotationally excited state is a factor 50 lower. We note that this method of generating an intense beam of slow, rotationally pure NO molecules may be of use in cold collision experiments \cite{Kirste2012,Gao2018}. 

We also observe a bright, continuous beam of cold NO molecules without ablation light, provided the second stage of the cryocooler is heated to above about 60\,K; below this temperature the signal completely disappeared. Spectra of the R$_1$/Q$_{12}(1/2)$ lines taken with the desorbed and continuous beams are shown in Figure \ref{fig:NO}c. Stick spectra shown pointing downwards are the individual hyperfine lines as predicted using the ground state parameters of \cite{Meerts1972} and the excited state parameters of \cite{Kaspar2022}. The simulated line positions and the experimental spectra are in good agreement. The fitted FWHM of the lines is larger by about a factor 1.7 for the continuous beam, from which we deduce a forward velocity of \SI{200}{\metre\per\second}. By comparison of the relative fluorescence signals, we find that the continuous beam produces the same number of molecules in the $J=1/2$ level in a \SI{5}{ms} time interval as a single desorption pulse. A spectrum for the R$_{1}/$Q$_{12}(3/2)$ line using the continuous source is also shown in Figure \ref{fig:NO}d. The probe laser power is equal for both spectra, showing that the $J=3/2$ level has greater population under these conditions. The experiments with NO suggest that a similar effect occurs for the fluorine donor gas during the production of metal fluoride molecules. The density of the reactant gas increases around the time of ablation by vaporisation of ice inside the cell. This may explain how reactant gas sources can be efficient despite the high activation energy of the reaction. 

\section{Results: Molecular Beams}

In this section, we determine the brightness of molecular beams of AlF, $^{40}$CaF, $^{24}$MgF and $^{174}$YbF and discuss the influence of experimental parameters on the molecular beam. For CaF, MgF and YbF, we probe using the Q$_1$(0) line of the A$^2\Pi_{1/2}, v'=0\leftarrow$X$^2\Sigma^+,v=0$ band. For these molecules the vibrational branching is negligible for short interaction times and we expect that at high laser intensities each molecule scatters on average $\mathcal{N} = 3$ photons before being optically pumped to $N=2$ in the vibronic ground state. The total emission pattern for the Q$_1(0)$ lines is isotropic. To probe the AlF beam, we use the A$^1\Pi, v'=0\leftarrow$X$^1\Sigma^+,v=0$ R($J$) lines, which optically pumps molecules from rotational state $J$ to $J+2$ in the vibronic ground state. For the R$(0)$ line, we also expect $\mathcal{N}=3$ photons per molecule in saturated fluorescence. The emission pattern has been measured to be slightly anisotropic by $\sim20\%$, so we use light which is linearly polarised parallel to the detector direction. In this configuration, the relative number of photons scattered on P($J$) and R($J$) lines has been experimentally found to be in agreement with the H\"{o}nl-London factors \cite{Hofsass2021}. 

\subsection{Comparison of Molecular Beams}

\begin{figure}
    \centering
    \includegraphics[width=0.5\textwidth]{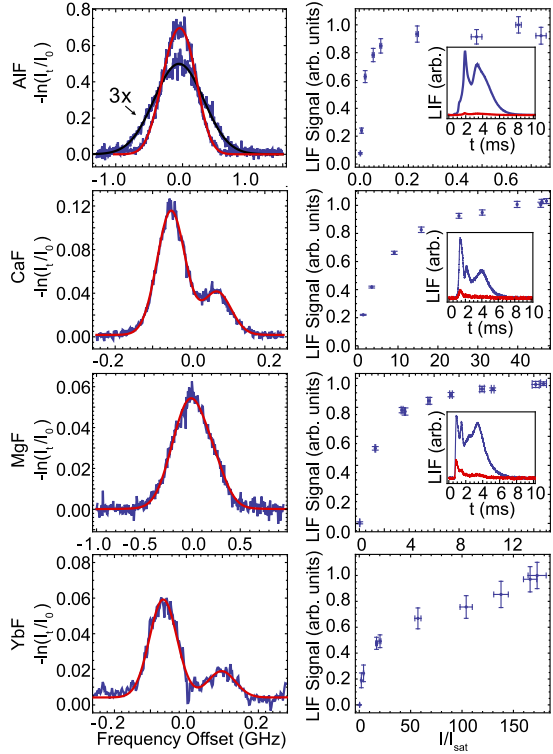}
    \caption{Left column: absorption spectra recorded directly outside the cell orifice. The solid red curves are fits based on the known energy structure of the molecules and using a Gaussian lineshape. The enlarged spectrum shown for AlF is measured \SI{20}{\milli\meter} downstream. Right column: saturation curves on the rotationally open transitions. For AlF, CaF and MgF, the insets show the fluorescence signal without (blue) and with (red) optical pumping applied in LIF-1, demonstrating that the molecules are indeed pumped away from the addressed transition. In the case of YbF, high rotational lines from other isotopologues prevent saturation.}
    \label{fig:source:Allmolecules}
\end{figure}

The left panels of Figure \ref{fig:source:Allmolecules} show the absorption spectra recorded directly after the buffer gas cell orifice for all four molecular species. The peak absorption of the AlF beam, presented in Figure \ref{fig:source:Allmolecules}a, is an order of magnitude higher than for the other monofluoride species. The absorption measured \SI{20}{\milli\meter} downstream is shown magnified in Figure \ref{fig:source:Allmolecules}a. This measurement reveals a broadening of the transverse velocity spread and is comparable to the atomic beams. We extract the transverse velocity widths by fitting the absorption spectra to Doppler-broadened line shapes, and use these to predict the brightness of each molecular beam.

The right panels in Figure \ref{fig:source:Allmolecules} show saturation of the fluorescence. The insets show the results of pump-probe measurements for AlF, CaF, and MgF demonstrating that the molecules are indeed optically pumped to $N=2$. For AlF, the large transition linewidth and small hyperfine structure in the $N=0$ ground state permits efficient saturation of the fluorescence at $I/I_{\textrm{sat}}<1$ and we observe a pumping efficiency of 97\%. For CaF, MgF and YbF a relatively high laser intensity of $I/I_{\textrm{sat}}>40$ is required to power-broaden the transition enough to cover the ground state spin-rotation and hyperfine structure. We verify this by fluorescence spectra at high power and pump-probe measurements. In the case of YbF, the background due to rotational lines of other molecular isotopologues leads to an increase in the fluorescence above $I/I_{\textrm{sat}}=100$ and we assume our conversion of fluorescence signal to molecule number is uncertain to within a factor two.

\begin{figure}
    \centering
    \includegraphics[width=0.5\textwidth]{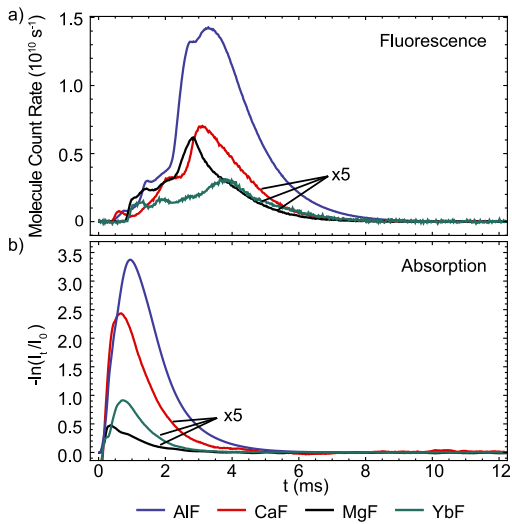}
    \caption{(a): Comparison of the molecule count rates in the ground rotational state for all species, obtained in saturated fluorescence, as a function of arrival time in the detector. We plot the count rate as the number of molecules entering the detector, limited by the $2\times2$\,mm$^2$ square aperture. (b): Comparison of the optical depth of the molecular beams, recorded at the buffer gas cell orifice. In both cases, the signal for MgF, CaF and YbF is magnified by a factor five.}
    \label{fig:source:LIFcomparison}
\end{figure}

Figure \ref{fig:source:LIFcomparison}a shows a comparison of the maximum fluorescence signals obtained --- under identical experimental conditions --- for all four molecules. The MgF, CaF and YbF signals are magnified by a factor 5. From these signals and the absorption measurements shown in Figure \ref{fig:source:LIFcomparison}b, we calculated the respective brightness in absorption and fluorescence. In Table \ref{table:source:atommoleculecomp}, we summarise our estimates of the brightness of each atomic and molecular beam source. For the molecules, the given values correspond to the number of molecules in the ground rotational state. In both absorption and fluorescence, AlF is the brightest molecular species by about a factor 10, while the brightnesses of the atomic beams are similar to within a factor two. This shows that the source output of the $^2\Sigma^+$ molecules is likely to be limited by the production of reaction by-products such as XF$_2$ \cite{Liu2022}, for $\textrm{X} = \textrm{Mg}$, Ca and Yb.  

\begin{figure}
    \centering
    \includegraphics[width=0.5\textwidth]{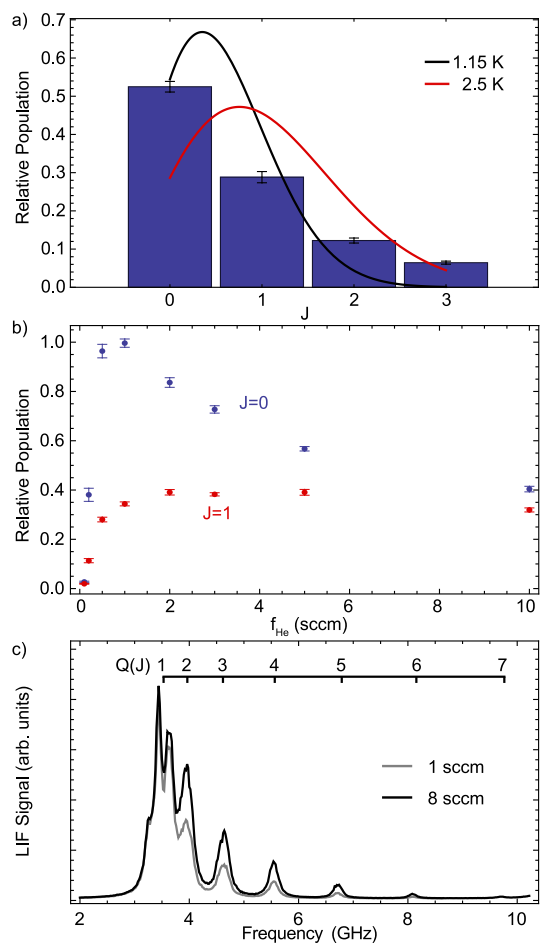}
    \caption{(a): Distribution of the AlF population over the four lowest rotational levels in the vibrational ground state of the \AlFgdstate state. The solid lines show Boltzmann distributions at $1.15$\,K and $2.5$\,K, respectively. (b): Influence of the helium buffer gas flow rate $f_{\mathrm{He}}$ on the fluorescence signal in LIF-2, detected on the R(0) line (blue) and the R(1) line (red). (c): Q-branch scans at He buffer gas flows of \SI{1}{sccm} and \SI{8}{sccm}.}
    \label{fig:source:Heflows}
\end{figure}

For AlF, we also recorded the relative rotational state populations by saturated fluorescence of successive R($J$) lines up to $J=3$. The resulting state distribution is presented in Figure \ref{fig:source:Heflows}a. It is poorly described by a thermal Boltzmann distribution, we show expected distributions for $T=1.15$\,K and $2.5$\,K as solid lines in the figure. The combined beam brightness for $0\leq J\leq 3$ is $3\times 10^{12}$ \SI{}{\per\steradian} per pulse, and consistent with the atomic aluminium beam measurement. This supports the idea that NF$_3$ reacts highly efficiently with aluminium atoms to produce AlF. Assuming a similar rotational distribution for CaF, we estimate a beam brightness which is $3\%$ of the atomic Ca beam. The helium flow has a significant effect on the rotational thermalisation in the beam, as illustrated in Figure \ref{fig:source:Heflows}b and c. The $J=0$ population is maximised at a He flow rate of about $\SI{1}{sccm}$, but it appears population is redistributed to higher rotational states as the flow rate increases. 

New simulation approaches for the in-cell dynamics will improve our understanding of the various influences on the velocity and internal state distributions of buffer gas molecular beams \cite{Schullian2015,Doppelbauer2017,Schullian2019,Takahashi2021,Schullian2022}. It has already been shown that the extraction efficiency can be enhanced for high buffer gas flow rates by choosing a specific cell geometry and by shaping the exit aperture \cite{Singh2018,Xiao2019}. 

The order of magnitude difference in molecule yield between AlF and the other monofluorides can be explained by the electronic structures of the atoms and molecules. Calcium and magnesium are group-II metals with a fully occupied $s$ orbital, ytterbium is a lanthanoid with fully occupied $f$ and $s$ orbitals, but aluminium is a group-III metal with an unpaired $p$ electron, making it a reactive radical. The gas-phase aluminium atoms react very strongly and efficiently with the fluorine donor gas, whereas for the other metal species, only a small percentage of the atoms reacts, as demonstrated in Figure \ref{fig:source:AlCaYbwithwithout}. While all studied monofluorides are stable molecules in the gas phase, only AlF possesses no unpaired electrons, making it much less reactive than the radicals CaF, MgF and YbF. The latter three are prone to the formation of difluorides. Theoretical calculations for CaF predict that a considerable amount of CaF$_2$, comparable to or more than the amount of CaF, is formed in the reaction of ablated calcium and a fluorine donor gas, while the difluoride production is suppressed for Al \cite{Liu2022}.

\begin{table} 
\caption{Properties of the atomic and molecular beams, as determined from fluorescence and absorption experiments. For the molecular species, we quote the brightness as observed in the ground rotational state. The transverse velocities $v_t$ are derived from the absorption spectra taken directly outside the cell exit aperture; where possible we state the transverse spread downstream of the cell in brackets. The combined uncertainties of the detection efficiencies lead to error bars for the LIF brightnesses of around 30\%. In the case of Yb and YbF, the given values are expected to be accurate within a factor 2 because we cannot use open transitions saturated by optical pumping.}
      \label{table:source:atommoleculecomp}
    \centering
\begin{tabular}{ p{2cm}|p{2cm}|p{4.cm}|p{4.cm} }
\hline
 Species & $v_t$(\SI{}{\metre\per\second}) & $\mathcal{B}_{\mathrm{f}}$ (sr$^{-1}$ per pulse)&  $\mathcal{B}_{\mathrm{abs}}$ (sr$^{-1}$ per pulse) \\
 \hline
 Al & 90 (150) & \AlAtomBrightnessLIF{} &  \AlAtomBrightnessAbs{} \\
 $^{40}$Ca  & 75 (120)& \CaAtomBrightnessLIF{} & \CaAtomBrightnessAbs{}\\
 $^{174}$Yb & 40 (60) & \YbAtomBrightnessLIF{} & \YbAtomBrightnessAbs{}\\
 AlF & 74 (160) & \AlFBrightnessLIF{} &  \AlFBrightnessAbs{} \\
 $^{40}$CaF & 53 (125)& \CaFBrightnessLIF{} & \CaFBrightnessAbs{}\\
 $^{24}$MgF & 112 & \MgFBrightnessLIF{} & \MgFBrightnessAbs{}  \\
 $^{174}$YbF & 50 & \YbFBrightnessLIF{} & \YbFBrightnessAbs{}\\
 
 \hline
\end{tabular}
\end{table}

\subsection{Velocity Distribution of the Molecular Beam}
\subsubsection*{Pump-Probe Method}
\begin{figure}
    \centering
    \includegraphics[width=\textwidth]{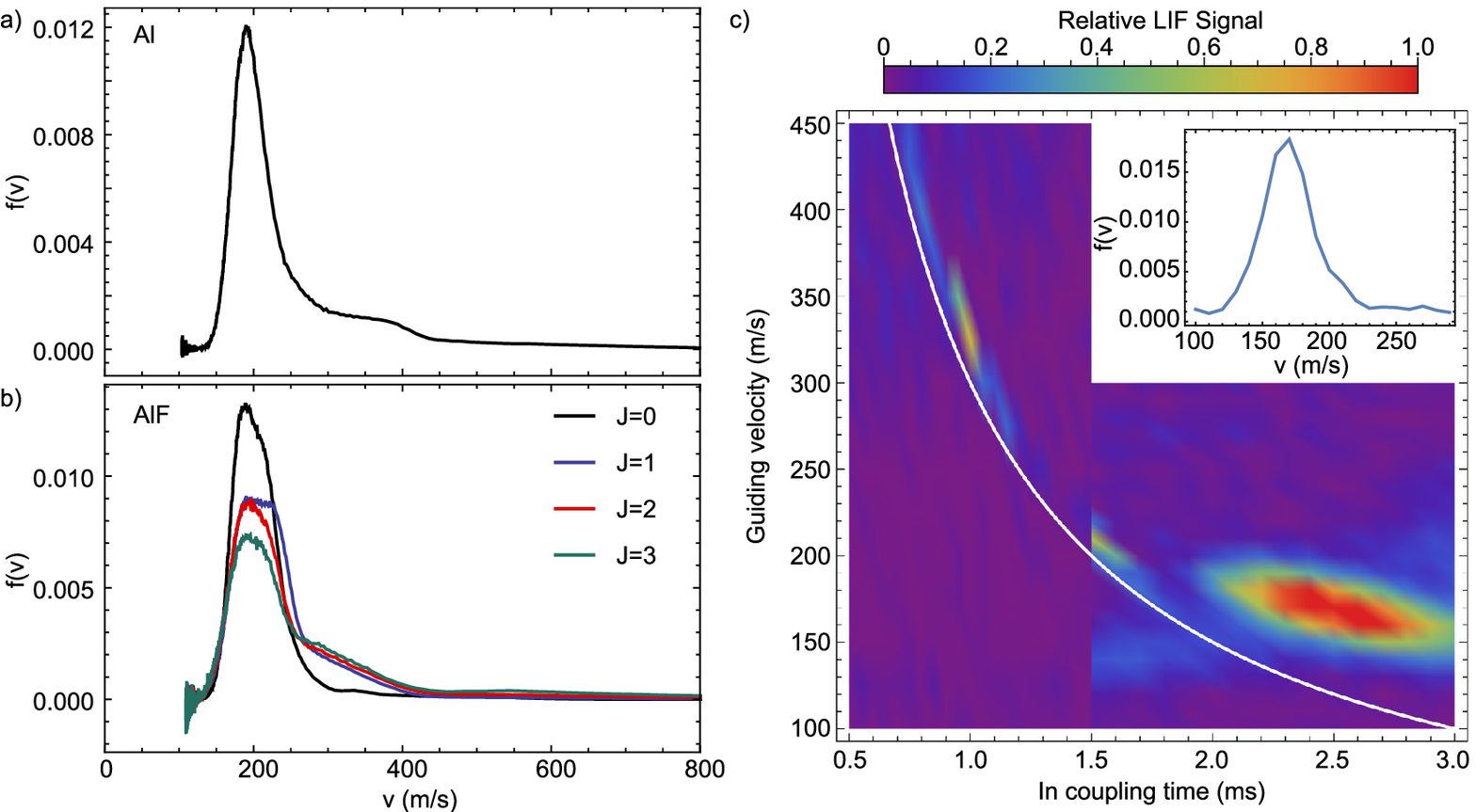}
    \caption{Velocity distributions, determined using a pump-probe method, of (a) the Al atomic beam and (b) the AlF molecular beams, for the four lowest rotational states in the \AlFgdstate{}, $v=0$ ground electronic state. (c): Phase space distribution of the AlF molecular beam, measured by velocity guiding $J=1$ molecules in a Stark Decelerator. The white curve shows the relation $v_z =z_0/t_i$ with $z_0 =\SI{300}{mm}$. The molecules which are well thermalised arrive with low velocities, and delayed by about \SI{0.7}{ms} relative to this ballistic trajectory. The inset shows the velocity distribution calculated by summing over all $t_i$ for $100<v_z<300$.}
    \label{fig:source:velocity}
\end{figure}

We determined the forward velocity distributions of the Al and AlF beams, using the pump-probe method described in \cite{Hofsass2021}. Briefly, applying optical pumping light in LIF-1 removes the signal in LIF-2; a short ($\sim\SI{10}{\micro\second}$) removal of the pump light results in an appearance of signal in LIF-2, in such a way that time of arrival in LIF-2 correlates with velocity. Repeatedly switching off the pump light at appropriate intervals as the molecules transit through LIF-1 then allows estimating the forward velocity distribution, $f(v)$, using a single molecular pulse with a high signal to noise ratio. We show the results of these experiments in Figure \ref{fig:source:velocity} for Al and AlF, where the velocity distributions for AlF are shown for $J=0-3$, and normalised such that the integral $\int f(v) dv =1$ for each $J$. The finite transit time through the laser beams corresponds to a velocity uncertainty of below $5\%$. Overall the velocity of the molecular and atomic beams are similar, and in the case of AlF, the ground rotational state contains the slowest molecules. 

\subsubsection*{Stark Decelerator}
To gain additional insight into the molecular buffer gas source, we replaced the LIF detection zones with a Stark decelerator to map out the longitudinal phase space distribution $(z,v_z)$ for AlF \cite{Bethlem1999}. The decelerator \cite{Engelhart2015} consists of 132 electrode pairs spaced at \SI{5.5}{\milli\meter} along the $z$-axis, whose orientation alternates between $\pm 45^{\circ}$ to the $x$- and $y$-axes in the $x-y$ plane. We operate the decelerator in the S=1 guiding mode (i.e. at a phase angle of zero), where the electrode configuration is switched synchronously with a molecule travelling at velocity $v_z$, which reaches the position of the decelerator entrance $z_0$ at a variable in-coupling time $t_i$. In this way, molecules in a small region of phase space ($z_0\pm\delta_z,v_z\pm\delta_v$) at time $t_i$ are guided through the machine and arrive at a defined time in the LIF detection zone, located at a short distance from the final electrode pair. The acceptance region $(\delta_z,\delta_v)$ is determined by the longitudinal electrode spacing, the peak electric field between the electrodes, and the number of used electrode stages. Along the $z$-axis, the peak electric field is about \SI{100}{kV\per\centi\metre}, corresponding to a potential depth of \SI{0.5}{K} for AlF molecules in weak field seeking states of the $J=1$ level. This results in $\delta_v = \SI{13}{\metre\per\second}$. 
 
Figure \ref{fig:source:velocity}c shows a false colour plot of the LIF signal as a function of the scanned parameters $t_i$ and $v_z$, for AlF molecules in the $J=1$ state. The white line in the figure marks the relation $v_z = z_0/t_i$ with $z_0 = \SI{30}{cm}$, representing ballistic motion from the target at the time of the ablation to the decelerator entrance. The buffer gas cooled pulse of molecules reaches the decelerator with an average velocity of about $\SI{170}{\metre\per\second}$, $\SI{0.7}{ms}$ later than the ballistic curve due to collisions and thus thermalisation with the buffer gas in the cell. A faster part of the beam arrives close to the ballistic trajectory curve near \SI{320}{\metre\per\second}, which presumably results from molecules which leave the cell shortly after the ablation without thermalising with the helium in the cell. This tends to occur over the course of several thousand shots necessary for such a measurement. In the inset of the figure, the velocity distribution is computed by integrating over all in-coupling times for $100<v<300$ \SI{}{\metre\per\second}. The distribution is narrower than measured by the pump-probe method in panel (b), with its peak velocity about \SI{40}{\metre\per\second} slower. The two different measurements were taken months apart, with the decelerator measurement using a freshly cleaned cell, and this can account for such a difference.

\subsection{Influence of the Ablation Laser Fluence}

\begin{figure}
    \centering
    \includegraphics[width=0.5\textwidth]{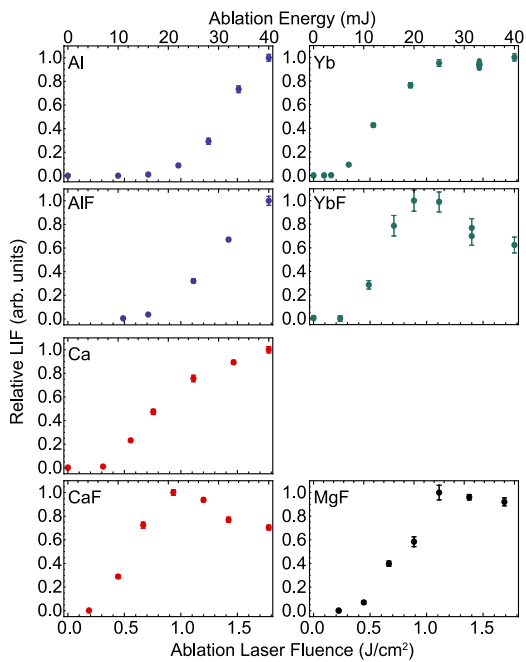}
    \caption{Fluorescence signals of all investigated species as a function of the ablation laser pulse energy (upper $x$-axes) and laser fluence (lower $x$-axes). We plot the relative LIF for the atomic beams and for the rotational ground state of the molecules.}
    \label{fig:source:yagpowers}
\end{figure}

The influence of the energy of the Nd:YAG infrared ablation laser on the number of atoms and molecules is shown in Figure \ref{fig:source:yagpowers}. The signal measured in LIF-2 is shown as a function of the ablation pulse energy and laser fluence using the measured spot size of \SI{0.7}{\milli\meter}. We observe that the threshold energy for observing Al and AlF is between 10 and \SI{15}{\milli\joule}, and the signal from both species increases up to the maximum available pulse energy of \SI{40}{\milli\joule}. In particular, the AlF yield is proportional to the ablation energy, and both would appear to benefit from increased pulse energy. For Ca and Yb, the threshold is $5-\SI{10}{\milli\joule}$, and we observe an optimum in respective molecular production at about \SI{25}{\milli\joule}. Above this pulse energy, the atomic signal either plateaus or weakly increases, but the number of molecules decreases. This may be due to incomplete thermalisation with the buffer gas, or changing reaction dynamics at higher ablation energies.

The differences in the phase explosion threshold energy for the metals used in our study likely arise from the difference in the latent heat of vaporisation of the atomic species, where the value for Al is roughly a factor two larger than for Ca, Mg and Yb. The reflectivity of the bulk metal surfaces at room temperature --- values for cryogenic temperatures are not reported --- is around 93\% for Al, Ca, and Mg, but 70\% for Yb \cite{LandoltBornstein1985}. The latter is a possible explanation for the large number of vaporised Yb atoms. 

\subsection{Influence of the Fluorine Donor Gas}

\begin{figure}
    \centering
    \includegraphics[width=0.5\textwidth]{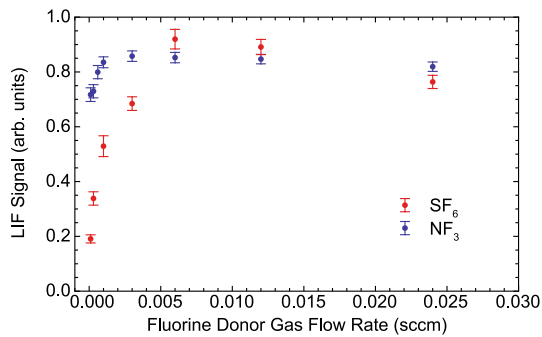}
    \caption{Influence of the fluorine donor gas flow rate on the MgF Q$_1$(0) fluorescence signal in LIF-2.}
    \label{fig:source:MgFfluorineflows}
\end{figure}

While the inert gas SF$_6$ is commonly used in experiments with monofluorides, theory suggests that the use of the more reactive and corrosive molecule NF$_3$ has advantages \cite{Liu2022}. We find that both produce a similar number of molecules when observed downstream of the source, within the range of day-to-day fluctuations of source operation with one fluorine donor gas. Overall, the response of molecule yield to the gas flow rates was not reliably reproducible from day to day, and strongly dependent on the flow rate history during a measurement sequence. We attribute this to probable freezing effect of the reactant gas discussed previously, and to the `poisoning' effect discussed below. However, we typically find NF$_3$ performs well with flow rates $\SI{0.001}{sccm}$ or below, but that SF$_6$ requires about ten times more flow to give comparable signal. This is illustrated in Figure \ref{fig:source:MgFfluorineflows} for the case of MgF. Operating the source with NF$_3$ is more reliable and results in a slower beam for a longer period of operation. SF$_6$ can produce a similar beam brightness with a freshly cleaned cell, but after one day of operation, the TOF profile shifts to higher velocities. This effect is particularly striking for AlF, where the forward velocity increases steadily with the number of ablation shots when SF$_6$ is used. We do not observe this effect when only flowing SF$_6$ or only ablating Al for a day. We assume that this speed-up is caused by the buildup of ablation products and sulphur-containing inorganic compounds on the internal buffer gas cell wall, thus hindering efficient (re-)thermalisation of the helium buffer gas. This effect is less pronounced for the other molecules. A larger internal volume of the cell is likely to improve this, but typically results in long temporal pulses. We find that extracting the molecular beam through the aperture into an extension that has the same bore diameter but no aperture reduces the forward velocity and significantly improves the long-term stability of the beam without affecting the beam brightness. This indeed indicates that the helium thermalisation is affected by ablation products in the main cell. An additional mesh on the extension reduces the mean forward velocity of AlF by up to \SI{70}{\meter\per\second}, but reduces the beam brightness \cite{Lu2011, Hofsass2021}.

\begin{figure}
    \centering
    \includegraphics[width=0.5\textwidth]{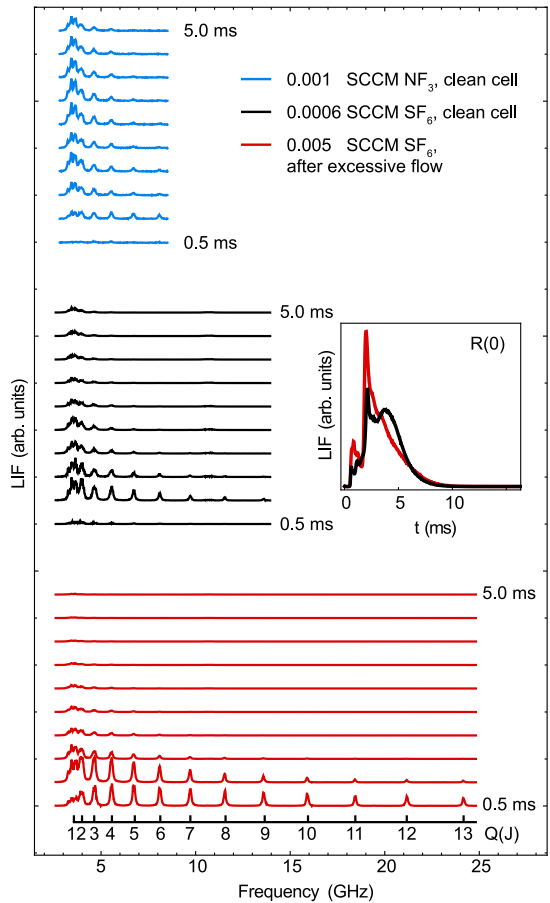}
    \caption{Q-branch spectra of the \mainAlFTrans transition in AlF selected by arrival time at the detector, in \SI{0.5}{ms} increments. The upper blue spectrum is recorded with a clean buffer gas cell using NF$_3$ gas (blue, top), whereas the lower two spectra are recorded with a clean (black, middle) and contaminated (red, bottom) cell using SF$_6$. The ruler inside the figure marks the different Q($J$)-lines. The inset shows the signal obtained for the $J=0$ population using the R($0$) line, for the two SF$_6$ examples.}
    \label{fig:source:poisonedsource}
\end{figure}

In Figure \ref{fig:source:poisonedsource}, we use the compact Q-branch of AlF as a convenient means to probe the rotational distribution reaching LIF-2, and thereby gain insight into the thermalisation dynamics in the cell. We plot a series of Q-branch spectra, separated by the arrival time of the molecules in the detection region, incremented in \SI{0.5}{\milli\second} steps. Thus, we can see the time dependent rotational distribution of the molecular beam. The upper blue set of spectra show typical behaviour when using NF$_3$ gas. The middle (black) set of spectra are measurements taken with a clean cell and an SF$_6$ flow rate of \SI{0.0006}{sccm}, and are similar to the NF$_3$ data. The red spectra are measurements with a flow rate of \SI{0.005}{sccm} after one day of source operation without cleaning afterwards. These spectra show that at a higher flow rate and in the used cell, the molecules arrive earlier at the detector, and are distributed over many rotational states; we indeed observe molecules in $J=13$ arriving \SI{1}{\milli\second} after the ablation pulse. The inset of the figure shows TOF fluorescence traces when saturating the R(0) line, demonstrating how the molecular pulse shifts towards higher velocities. We find that the onset of this reduced thermalisation occurs more rapidly as the flow rate increases, and that good thermalisation can only be recovered by cleaning ablation products from the buffer gas cell. The rotational spectra for different molecule arrival times also serve to show that translational and rotational energy of the AlF molecules are correlated - fast molecules are rotationally hotter than slow ones.

Sulphur hexafluoride and nitrogen trifluoride are not the only possible fluorine donors. We performed preliminary experiments using tetrafluoromethane (CF$_4$) as fluorine donor gas that lead to similar molecule yields as for NF$_3$ and SF$_6$, and we therefore did not further pursue this avenue. Solid xenon difluoride (XeF$_2$), theoretically predicted to be a good candidate fluorine donor molecule \cite{Liu2022}, did not prove as a viable option. At room temperature, the compound evaporates quickly into a hazardous gas. Nevertheless, we connected a reservoir containing XeF$_2$ crystals to the buffer gas cell and adjusted the flow with a needle valve. We observed efficient molecule production, but the relatively high flow rate resulted in a very fast molecular beam.

\section{Conclusion and Outlook}

We have presented a series of experiments comparing buffer gas cooled beams of Al, Yb, Ca, and their monofluorides AlF, CaF, MgF and YbF formed by reaction of laser-ablated atoms with a fluorine donor gas inside a buffer gas cell. We find that the molecular beam brightness of AlF is about one order of magnitude larger than for the other monofluorides, and when multiple rotational states are considered we observe a similar beam brightness to the atomic Al beam. This is qualitatively consistent with the near complete loss of atomic signal upon introducing the reactant gas into the cell. In contrast, the Ca and Yb atomic beams are relatively unaffected by the fluorine reagents, and brighter than the Al beam. The molecular yield of CaF, MgF and YbF suggests a reaction efficiency on the few percent level. This difference in reactivity is explained by the radical character of aluminium and the stability of the AlF molecule, while the other molecules are radicals that are formed from less reactive atoms.

We demonstrated high-flux pulsed and continuous molecular beams of buffer gas cooled NO, and performed CW UV spectroscopy of low-$J$ lines of the A$^2\Sigma^+,v'=0\leftarrow$ X$^2\Pi_{1/2},v=0$ transition. The low forward velocity and rotational state purity of the pulsed beam provide a useful starting point for low-energy collision experiments with NO. Our observations also provide insight about the vaporisation of ice in the cell, and the pulsed desorption method may be applicable to other diatomic molecules (e.g. H$_2$, O$_2$).

We conclude with some guidance for others wishing to develop cryogenic buffer gas beam sources. First, we find it useful to measure the molecular beam brightness with both absorption and fluorescence methods, noting that absorption directly outside the cell leads to an overestimate. A large discrepancy between the two measurements indicates insufficient cryopumping or deteriorating charcoal. Second, monitoring the atomic buffer gas beams, especially in sources using reactant gases, provides a useful reference when studying the molecular beam properties. Finally, thermalisation inside the cell is influenced by both the choice of reactant gas, and the condition of the internal cell surfaces. NF$_3$ reacts more efficiently to form fluoride molecules as a significantly lower flow rate is required to produce the same molecular beam brightness. The lower freezing point also allows cooling the capillary that feeds the fluorine donor gas, which reduces the thermal heat load on the cell. We also find that the molecular beam parameters, especially for AlF, are more consistent when using NF$_3$. 

\section*{Acknowledgements}
We thank Sebastian Kray and Klaus-Peter Vogelgesang as well as the mechanical workshop of the Fritz Haber Institute for expert technical assistance. We are grateful to Jes\'{u}s P\'{e}rez R\'{i}os for helpful and stimulating discussions which prompted this study. This project received funding from the European Research Council (ERC) under the European Union’s Horizon 2020 Research and Innovation Programme (CoMoFun, Grant Agreement No. 949119).

\section*{Data Availability}
The data presented can be accessed from Zenodo (\url{https://doi.org/10.5281/zenodo.7063976}) and may be used under the Creative Commons Attribution 4.0 International license.

\bibliographystyle{tfo}
\bibliography{sourcepaper}

\end{document}